\documentclass[11pt]{article}
\usepackage{moriond}
\usepackage{slashed}
\usepackage[pdftex]{graphicx}
\pdfoutput=1

\def\Journal#1#2#3#4{{#1} {\bf #2}, #3 (#4)}

\def\PLB{{\em Phys. Lett.}  B}
\def\PRL{\em Phys. Rev. Lett.}
\def\PRD{{\em Phys. Rev.} D}

\def\ra{\rightarrow}

\def\be{\begin{equation}}
\def\ee{\end{equation}}
\def\bea{\begin{eqnarray}}
\def\eea{\end{eqnarray}}

\begin{document}
\vspace*{4cm}
\title{HIGGS TO FOUR TAUS AT ALEPH}

\author{ JAMES BEACHAM, \\ on behalf of the ALEPH Collaboration \\ \quad }

\address{Department of Physics, New York University, 4 Washington Place,\\
New York, NY, USA}

\maketitle\abstracts{
A search has been performed on 683 pb$^{-1}$ of data collected by the ALEPH detector at the Large Electron-Positron (LEP), collider at centre-of-mass energies from 183 to 209 $\mathrm{GeV}$ looking for a Higgs boson decaying into four $\tau$ leptons via intermediate pseudoscalar $a$ particles, for a Higgs mass range of 70 to 114 $\mathrm{GeV} / c^2$ and an $a$ mass range of 4 to 12 $\mathrm{GeV} / c^2$.  No excess above background is seen and a limit is placed on $\xi^2 = \frac{\sigma(e^+ e^-\ra Z+h)}{\sigma_{SM}(e^+ e^-\ra Z+h)}\times(h\ra aa)\times(a\ra \tau^+\tau^-)^2$ in the $m_h, m_a$ plane.  For $m_h < 107 \; \mathrm{GeV} / c^2$ and $m_a < 10 \; \mathrm{GeV} / c^2$, $\xi^2 > 1$ can be excluded at the 95\% confidence level.
}

\section{Introduction}

Direct searches at LEP2 for the standard model (SM) Higgs boson, $h$, decaying into $b \bar{b}$ or $\tau^+ \tau^-$ placed a lower bound of 114 $\mathrm{GeV} / c^2$ on the Higgs mass~\cite{lephiggs}.  However, fits of the SM to electroweak precision data suggest a Higgs with a mass within the kinematic limit of LEP.  Additionally, a small, non-SM-like excess observed at a Higgs mass of around $100 \; \mathrm{GeV} / c^2$ in the $b \bar{b}$ final state at LEP and the fine-tuning needed in the minimal supersymmetric standard model (MSSM) have led to the consideration of models, such as the next-to minimal supersymmetric standard model (NMSSM)~\cite{derm2}, that feature exotic Higgs boson decays and naturally light pseudoscalar particles, $a$.  In these models, new decay channels, such as $h \ra aa$, can dominate over $h \ra b \bar{b}$ and render the Higgs boson ``invisible" to conventional searches.   In particular, the Higgs can decay into four SM particles instead of two, via two intermediate $a$ particles.  Several of these possible final states, such as $h \rightarrow 2a \rightarrow 4 b$, are already highly constrained by existing analyses; see Ref.~\cite{lephiggs2}, for example.  For $m_a < 2 m_b$, however, the decay $a \ra \tau^+ \tau^-$ is expected; this process, with the Higgs decaying into $\tau^+\tau^-\tau^+\tau^-$ for a Higgs mass range of 86 to 114 $\mathrm{GeV} / c^2$, is not covered by existing analyses.  To investigate this range, the ALEPH data has been revisited.  The present analysis is described in detail in Ref.~\cite{al}.

\section{The ALEPH Detector}

A detailed description of the ALEPH detector can be found in Ref.~\cite{aldet} and of its performance in Ref.~\cite{alper}.  High momentum resolution is achieved via a large tracking volume immersed in a 1.5 T magnetic field.  An energy-flow reconstruction algorithm measures total visible energy in the event by combining measurements from the tracking sub-detectors and the electromagnetic and hadronic calorimeters, and provides a list of reconstructed objects (\emph{energy-flow objects}) which are classified as charged particles (which correspond to charged particle tracks, here called \emph{tracks}), photons, and neutral hadrons.  These energy-flow objects are the basic entities used in the present analysis.

During LEP2 the machine operated at centre-of-mass energies from 183 to 209 $\mathrm{GeV}$ and collected data corresponding to a total integrated luminosity of 683 pb$^{-1}$.

\section{Signal and Background Samples}

We revived all steps of the ALEPH analysis framework, including the ability to generate simulated samples of standard model background and data.  We produced 3000 simulated signal events (with $h \rightarrow aa$ followed by $a \rightarrow \tau^+\tau^-$) for each of the three Z decay channels considered and for each combination of Higgs boson and pseudoscalar masses in the ranges 70 $<$ $m_h$ $<$ 114 $\mathrm{GeV} / c^2$ and 4 $<$ $m_a$ $<$ 12 $\mathrm{GeV} / c^2$ in steps of 2 $\mathrm{GeV} / c^2$.  For the relevant background processes, our samples were either 10-30 or 300-1000 times larger than the data, depending upon the process.

\section{Event Selection}

For the mass range considered, the Higgs is produced approximately at rest, and thus the decay $h \rightarrow 2a \rightarrow 4 \tau$ results in a pair of taus recoiling against another pair of taus.  For the $a$ mass range considered, the decay products of each 2$\tau$ system will be observed as a highly-collimated jet of charged particles.  Due to this high level of collimation, individual identification of taus, via standard algorithms, would fail.  Instead, we used the fact that each tau decays into either one charged particle or three charged particles, and we would thus expect each $a$ jet to contain two, four or six tracks.  We used the $\mathrm{JADE}$ algorithm to form jets with a $y_{cut}$ chosen to merge proto-jets up to a mass of $m_{jet} =$ 15 $\mathrm{GeV} / c^2$.

We considered three possible decays of the Z boson, namely Z $\rightarrow e^+e^-$, Z $\rightarrow \mu^+\mu^-$, and Z $\rightarrow \nu\bar{\nu}$, and formulated a set of \emph{loose} selection criteria (convenient to allow comparison of data and simulation at an intermediate stage without compromising the blind nature of the analysis) and \emph{final} selection criteria for each of the two Z decay classes considered: Z $\rightarrow l^+l^-$ (where $l = e$ or $\mu$) and Z $\rightarrow \nu\bar{\nu}$.

\subsection{Z $\rightarrow l^+l^-$}

For the Z $\rightarrow l^+l^-$ channel, four-fermion background processes are prominent.  We used ALEPH lepton identification algorithms to mask the two most energetic leptons in the event from the list of objects clustered by the $\mathrm{JADE}$ jet-finding algorithm.  The loose selection consisted of the following requirements: Two oppositely-charged, isolated leptons; two jets, well-contained within the tracking volume ($| \cos \theta_j | < $ 0.9 ); and the jets and leptons sufficiently isolated from each other ($| \cos \theta_{jl}^{min} | < $ 0.95).  The final selection consisted of the following requirements: The invariant mass of the lepton pair near the Z mass (80 $< m_{l^+l^- (\gamma)} <$ 120 $\mathrm{GeV} / c^2$, where $\gamma$ indicates an isolated photon that may have been radiated from one of the leptons and which is added to the di-lepton system if doing so corresponds to an invariant mass closer to the Z mass than the di-lepton pair alone); missing energy due to neutrinos from tau decays ($\slashed{E} >$ 20 $\mathrm{GeV}$); jets sufficiently separated ($| \cos \theta_{jj} | < $ 0); and a signal-like track multiplicity, i.e., each jet must contain either two or four tracks.

\subsection{Z $\rightarrow \nu\bar{\nu}$}

The Z $\rightarrow \nu\bar{\nu}$ channel represents a larger branching ratio of the Z than the lepton channel, and thus drives the analysis.  A major background contribution arises from $\gamma \gamma$ events.  The loose selection consisted of the following requirements: Modest missing energy and missing mass ($\slashed{E} >$ 30 $\mathrm{GeV}$ and $\slashed{m} >$ 20 $\mathrm{GeV} / c^2$); exactly two jets, well-contained in the tracking volume, with a modest invariant mass cut on the dijet system ($| \cos \theta_j | < $ 0.85 and $m_{jj} >$ 10 $\mathrm{GeV} / c^2$); requirements on the angle of the missing momentum vector with the beam axis and the total visible energy in the event, to reject substantial portions of two-photon-initiated and beam background events ($| \cos \theta_{miss} | < $ 0.9 and $E_{vis} >$ 0.05 $E_{CM}$); and modest requirements on the most energetic jet ($E_{j_{1}} >$ 25 $\mathrm{GeV}$ and containing either two or four tracks).  The final selection consisted of the following requirements: Less than 5 $\mathrm{GeV}$ within 30$^{\circ}$ of the beam axis, to reject events with energy deposits in the forward region of the detector; consistency with the Z boson decaying to neutrinos ($\slashed{E} >$ 60 $\mathrm{GeV}$ and $\slashed{m} >$ 90 $\mathrm{GeV} / c^2$); small aplanarity ($<$ 0.05), consistent with two back-to-back, highly collimated jets; and a signal-like track multiplicity, i.e., each jet must contain either two or four tracks.

\section{Results}

Based upon these selection criteria, our signal efficiency ranged from $\sim$25$\%$ to $\sim$50$\%$, depending on Z decay channel, Higgs mass, and $a$ mass.  We determined that, for the Z $\rightarrow l^+l^-$ channel, we should expect $\sim$3 signal events versus $<$ 0.2 background events, and for the Z $\rightarrow \nu\bar{\nu}$ channel our expectation was $\sim$11 signal events versus $\sim$6 background events.

Systematic uncertainties in our Monte Carlo simulation were estimated to be 5$\%$ for all signal and 10$\%$ for background in the Z $\rightarrow l^+l^-$ channel versus 30$\%$ for background in the $\textrm{Z} \rightarrow \nu\bar{\nu}$ channel.  We found that the background estimate and the number of events seen in data at the loose selection agreed within the systematic and statistical uncertainty for all Z channels.

For the Z $\rightarrow l^+l^-$ channels, we observed zero events after applying all selection criteria, while for the Z $\rightarrow \nu\bar{\nu}$ channel we observed two events.  These observations are consistent with background.

\begin{figure}
\begin{center}
\includegraphics[scale=0.53]{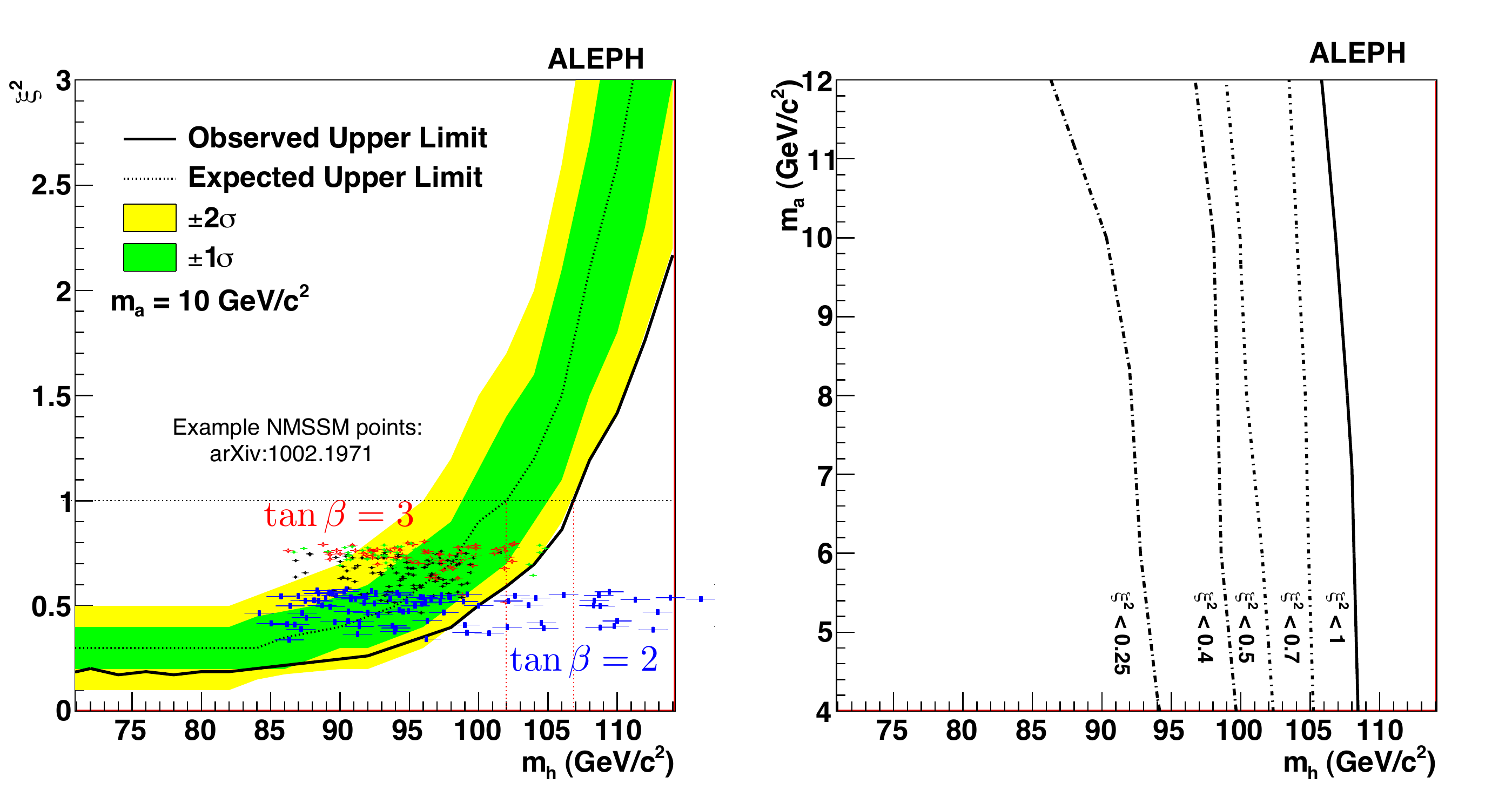}
\end{center}
\caption{Left: Observed and expected 95\% CL limit on $\xi^2$ as a function of $m_h$ for $m_a = 10 \; \mathrm{GeV} / c^2$.  Also shown are some favored points in the parameter space of the NMSSM.  Right: Contours of observed 95\% CL limit on $\xi^2$ in the $m_h, m_a$ plane.
\label{fig:results}}
\end{figure}

We place limits on the cross section times branching ratio of our signal process with respect to the SM Higgsstrahlung production cross section,  $\xi^2 = \frac{\sigma(e^+ e^-\ra Z+h)}{\sigma_{SM}(e^+ e^-\ra Z+h)}\times(h\ra aa)\times(a\ra \tau^+\tau^-)^2$.  The limits are based upon event counts in three separate track multiplicity bins (corresponding to events with two jets where 1) each jet contains two tracks, 2) each jet contains four tracks, or 3) one jet contains two tracks while the other contains four tracks) times each of the three Z decay channels considered, resulting in nine categories.  The resulting joint probability density for the event counts is then used to construct confidence intervals using a generalized version of the Feldman-Cousins technique~\cite{fc}, which incorporates systematic uncertainties in a frequentist way~\cite{cl}~\cite{cran}.  Results are shown, for the 95$\%$ confidence level, as a function of $m_h$ (for $m_a =$ 10 $\mathrm{GeV} / c^2$) on the left in Fig.~\ref{fig:results} and as contours within the $m_h, m_a$ plane on the right in Fig.~\ref{fig:results}.  Note that our selection criteria do not depend on $m_h$ or $m_a$, and thus our upper limits are fully correlated.  The observed number of events is consistent with a downward fluctuation of the background and, as such, our limits on $\xi^2$ are stronger than expected.

Also shown on the left in Fig.~\ref{fig:results} is the effect of these results upon some possible favored scenarios in the NMSSM; see Ref.~\cite{derm}, Figures 17 and 21 therein.  Our limits highly constrain scenarios with $\tan \beta \ge$ 3, while scenarios with $\tan \beta \le$ 2, where there is a larger branching ratio of the Z boson into jets, remain unconstrained.

\newpage

\section{Conclusions}

We have performed a search for a Higgs decaying into four taus via Higgstrahlung at LEP2, for the process $h \rightarrow 2a \rightarrow 4\tau$ and Z $\rightarrow e^+e^-$, $\mu^+\mu^-$, or $\nu\bar{\nu}$, using ALEPH data.  We observed no excess above background, and for $m_h < 107 \; \mathrm{GeV} / c^2$ and $m_a < 10 \; \mathrm{GeV} / c^2$, $\xi^2 > 1$ can be excluded at the 95\% CL, where $\xi^2 = \frac{\sigma(e^+ e^-\ra Z+h)}{\sigma_{SM}(e^+ e^-\ra Z+h)}\times(h\ra aa)\times(a\ra \tau^+\tau^-)^2$.  This analysis covers a region of parameter space previously unexplored and further constrains models that feature light pseudoscalar Higgs particles and non-standard Higgs decays, such as the NMSSM.

\section*{Acknowledgments}
The author would like to thank K. Cranmer, I. Yavin and P. Spagnolo for their fruitful collaboration, as well as all members of the ALEPH collaboration for the successful design, construction and operation of the ALEPH detector.

\end{document}